\documentclass[12pt]{emulateapj}
\newcommand{\beq}{\begin{equation}}
\newcommand{\eeq}{\end{equation}}

\newcommand{\hi}{H{\sc i}~}

\newcommand{\citei}[1]{\citeauthor{#1} \citeyear{#1}}

\usepackage{color}
\slugcomment{Accepted to the ApJ}

\begin{document}

\title{A Correction to the Standard Galactic Reddening map: Passive Galaxies as Standard Crayons}

\author{J.~E.~G.~Peek\altaffilmark{1}}
\author{Genevieve J.~Graves\altaffilmark{2}}
\altaffiltext{1}{Department of Astronomy, Columbia University, New York, NY. goldston@gmail.com}
\altaffiltext{2}{Department of Astronomy, University of California Berkeley, Berkeley, CA 94720}

\begin{abstract}
We present corrections to the \citeauthor{SFD98}~(\citeyear{SFD98};
SFD98) reddening maps over the Sloan Digital Sky Survey northern
Galactic cap area. To find these corrections, we employ what we dub
the ``standard crayon'' method, in which we use passively evolving
galaxies as color standards by which to measure deviations from the
reddening map. We select these passively evolving galaxies
spectroscopically, using limits on the H$\alpha$ and [O\textsc{ii}]
equivalent widths to remove all star-forming galaxies from the SDSS
main galaxy catalog. We find that by correcting for
known reddening, redshift, color-magnitude relation,
and variation of color with environmental density, we can reduce the
scatter in color to below 3\% in the bulk of the 151,637 galaxies we
select. Using these galaxies we construct maps of the deviation from
the SFD98 reddening map at 4.5$^\circ$ resolution, with 1-$\sigma$
error of $\sim$ 1.5 millimagnitudes $E(B-V)$. We find that the SFD98
maps are largely accurate with most of the map having deviations below
3 millimagnitudes $E(B-V)$, though some regions do deviate from SFD98
by as much as 50\%. The maximum deviation found is 45 millimagnitudes
in $E(B-V)$, and spatial structure of the deviation is strongly
correlated with the observed dust temperature, such
  that SFD98 underpredicts reddening in regions of low dust
  temperature.  Our maps of these deviations, as well as their errors, are made available to
the scientific community as supplemental correction to SFD98 at \url{http://www.peekandgraves2010.com}.
\end{abstract}

\keywords{astronomical databases: atlases,  ISM: dust, extinction, Galaxy: local interstellar matter, galaxies: photometry, galaxies: elliptical and lenticular, cD, Infrared: ISM}

\section{Introduction}\label{intro}
The Galaxy's interstellar medium (ISM) is suffused with particulate
material called interstellar dust. This dust, produced by stars and in dense ISM regions, absorbs and scatters (extincts) light and therefore contaminates our detections of objects beyond the Galaxy. In particular, this extinction is a strong function of the frequency of the light, eliminating blue and ultraviolet light much more than red and infrared light. Without an accurate map of these reddening and extinction effects, observations of distant objects have incorrectly measured colors and brightnesses. 

Historically, two methods were used to determine the extinction by
Galactic dust. Galaxy counts were used by \citet{SW67}, under the assumption that fewer galaxies would be found behind regions with more Galactic dust. \citet{KK74} used \hi column density, making use of the known relationship between \hi column and dust column. \citet{BH78} elegantly merged these two techniques to make up for each of their weaknesses; the galaxy count method is susceptible to angular-density fluctuations caused by the large-scale structure of galaxies on the sky (LSS) and the \hi method is susceptible to variations in the dust-to-gas ratio. 

The release of the \citeauthor{SFD98} (\citeyear{SFD98}; SFD98) reddening maps represented a further, and dramatic, step forward in our knowledge of Galactic extinction. The method they employed depends upon the fact that Galactic dust radiates infrared emission and is generally optically thin at those wavelengths. Therefore, with an infrared emission map and a dust temperature map one can deduce the dust column, if assumptions are made about the radiative efficiency of the dust.  
For the infrared maps they generated a 100 micron infrared map at 6.1$^\prime$ resolution from the combination of the \emph{Infrared Astronomy Satellite (IRAS)} data and the DIRBE (Diffuse Infrared Background Experiment) instrument on the \emph{Cosmic Background Explorer (COBE)} . To determine the temperature they made 100 and 240 micron maps at 1.3$^\circ$ resolution generated from DIRBE. To use the notation of SFD98,
\beq\label{expo}
E\left(B-V\right) = p I_{corr} X\left(R \right),
\eeq
where $E\left(B-V\right)$ is the standard selective extinction measure,
$I_{corr}$ is the 100 micron map (corrected for residual striation, zodical light, and the uniform extragalactic background from distant galaxies), X is a derived coefficient
such that $I_{corr}X$ represents a temperature-corrected infrared
flux, $R$ is the ratio of the 100 micron flux to the 240 micron
flux, and $p$ is a constant.

Twelve years after the publication of SFD98, we wish to put these
extinction measurements through a rigorous test using a principle we
call the ``standard crayon.'' In analogy to the standard candle
principle, wherein the observed brightness of an
  object of known luminosity allows one to determine a distance, a
standard crayon is an object of known intrinsic \emph{color} whose
observed color allows one to determine the reddening of the
intervening material. We note that SFD98 has been tested in the past
(e.g. \citei{AG99}, \citei{Chen99}, \citei{Stanek99},
\citei{Cambresy01}, \citei{YFS07}, and \citei{RF09}), but that these
works typically examine the reddening correction over small areas or
at relatively high extinction. Our analysis is the first to constrain
(and correct) errors in the SFD98 map over large areas of sky at high
precision. We also acknowledge the excellent work of Schlafly et al\. , who have been conducting similar tests to our
own using the blue tip of the stellar locus, and whose work is
currently in preparation. Our results are largely in agreement with
their work, though each work concentrates on different aspects of the
reddening correction and its errors. In particular, their data cover areas with higher extinction, and as such they are able to address questions of variability in the Galactic extinction curve.

In this work we use passively evolving (quiescent) galaxies as our standard crayons,
although in principle any object beyond the Galactic dust with a known
color could be used to do a similar experiment. 
  Galaxies are well suited to this task for several reasons.  They lie
  outside the Milky Way and thus probe the entire column of Galactic
  dust along the line of sight.  They are numerous and tile the sky
  relatively densely.  And most critically, their intrinsic colors can
  be predicted based on other photometric and spectroscopic galaxy
  properties, making it possible to measure deviations between their
  observed colors and their predicted intrinsic colors. 

These passively evolving galaxies, whose star formation has ceased, are observed to
  populate a tight sequence in the color-magnitude diagram that is
  quite cleanly separated from the population of blue star-forming
  galaxies (e.g., \citealt{strateva01, hogg03, balogh04, baldry04}).
  Furthermore, this ``red sequence'' is tilted such that the intrinsic
  colors of the galaxies are correlated with their absolute
  magnitudes.  Thus, the magnitudes of red sequence galaxies can be
  used to determine their expected colors.  Another practical advantage
  of red sequence galaxies is that they have been surveyed in large
  numbers using consistent observational methods, which minimizes the
  effect of systematic errors on relative galaxy colors.  In
  particular, this work makes use of the Sloan Digital Sky Survey
  (SDSS) \citep{york00}, which provides an enormous extant survey of
  galaxies that have been observed in a uniform manner.  

To implement our method we must determine the residual variation in passive galaxy color that could be attributable to errors in the SFD98 reddening map. The first step is to move all the galaxies to a standard reference frame using K-corrections and to remove the effect of extinction from dust using SFD98. After this, we determine the color variation of these galaxies with intrinsic properties such as redshift, absolute magnitude, and galactic environment, so that we can correct for these variations. We also attempt to determine variations in galaxy color that depend on observing parameters, such as errors in the SDSS photometry (see \S \ref{pc}). After applying these corrections, we are left with a color residual for each galaxy, which will be a combination of errors in the SFD98 reddening map, intrinsic scatter in galaxy color, and measurement noise. Since the intrinsic scatter in galaxy color dominates the effect from errors in the reddening map, we must average over nearby galaxies to determine a typical error for a region of the sky and generate a map of errors in the SFD98 reddening correction map.

In \S \ref{gs} we describe the SDSS observations and how we select our subsample of quiescent galaxies. In \S \ref{sc} we explain how we correct these galaxy colors to use them as standard crayons, as well as describe the resulting color distribution. In \S \ref{ecm} we describe our map-making methodology and explore various possible sources of error in the maps. In \S \ref{results} we show our final maps and results. We conclude in \S \ref{conc}.

\section{Galaxy Sample}\label{gs}

There are a number of considerations that determine
  the properties of the ideal standard crayon galaxies.

First, we require a set of galaxies that possess an
  intrinsically tight color-magnitude relation (CMR).  Substantial
  scatter in the intrinsic relation has a couple of negative effects:
  it reduces our ability to accurately constrain the mean
  color-magnitude relation, and it lowers the signal-to-noise ($S/N$)
  ratio of any detected color residuals.  
  This requirement has to be offset against the need to
  include enough galaxies to have reasonable statistics, as the signal
  we are trying to measure is weak.

Intrinsic scatter and random noise can in theory be
  beaten down with large numbers.  More problematic are
  spatially correlated color variations, since these will masquerade
  as the variations in the foreground reddening we are trying to
  measure.  There are four catagories of color variations
  that could potentially cause problems: intrinsic color variations in the
  population, variations in the selection bias of the sample,
  variations in photometry errors, and variations in spectroscopic
  errors.\footnote{These last are relevant because we use spectroscopic
    criteria to select passively-evolving galaxies (see section
    \ref{selection_criteria}).} If any of these four types of
  variations are spatially correlated, they will create spurious
  signal in the derived reddening maps.

Galaxies, and red sequence galaxies in particular, are
  known to be strongly clustered in three-dimensional space.  In
  addition to potentially creating spatially varying signals that are
  projected onto the sky, this correlation also means that redshift-dependent
  variations (such as errors in the K-correction, or problems with
  detecting emission in the presence of strong skylines) can result in
  spatially correlated color variations.

With these considerations in mind, this section
  describes the galaxy sample selection in detail.  In section
  \ref{sc}, we describe further corrections to the sample in order to
  account for spurious sources of color variation that can contaminate
  our reddening maps.

\subsection{SDSS Observations}

The galaxies used in this analysis are from the SDSS
  spectroscopic Main Galaxy Survey \citep{strauss02} Data Release 7
  (DR7, \citealt{adelman-mccarthy08}), which is an $r$-band magnitude-limited sample with no color selection.  We do not use galaxies from
  the SDSS Luminous Red Galaxy sample \citep{eisenstein01}, since
  these galaxies were targeted based on color cuts and therefore may have
  substantial color biases.  Photometric and spectroscopic parameters
  were downloaded from the NYU Value Added Catalog (VAGC,
  \citealt{blanton05-vagc}), which uses the updated SDSS photometric
  calibration described in \citet{padmanabhan08}. (We follow the colloquial terminology and refer to this work as ``\"ubercal''). We note that \"ubercal is optimized for sources with $r$-$i$ color of 0.2, and our median galaxy $r$-$i$ color is 0.4, which may introduce some very small systematic error. We use the K-corrections from \citet{BR07} version 4.2; earlier versions of this K-correction algorithm had problems in the bandpass projection which caused detectable systematics in our results.  These data were
  supplemented with emission line measurements and spectral models
  from the MPA-JHU release of DR7 spectrum
  measurements\footnote{\url{http://www.mpa-garching.mpg.de/SDSS/DR7/}}.  We
  also make use of the projected 5th-nearest-neighbor over-density
  ($\delta_5$, \citealt{cooper08}) to determine the local
  environment of each galaxy in DR7 (M.\ Cooper, private
  communication). Cross-matching the NYU VAGC, MPA/JHU spectroscopic measurement
catalog, and the \citet{cooper08} environment catalog produces a
parent sample of 665,107 galaxies that were observed as part of the
Main Galaxy Sample \citep{strauss02}.  

\subsection{Galaxy Selection Criteria}\label{selection_criteria}

For this work, we wish to select a sample of
  passively evolving red sequence galaxies for our set of standard
  crayons {\it using criteria that do not include their observed
    colors.} Any selection based on galaxy color could end up selecting for over- or under-reddened galaxies, biasing the sample.  Instead, we choose galaxies to be
  quiescent, non--star-forming galaxies by requiring that their SDSS
  spectra are free of detectable emission lines.  

\begin{figure}
\begin{center}
\includegraphics[scale=0.9]{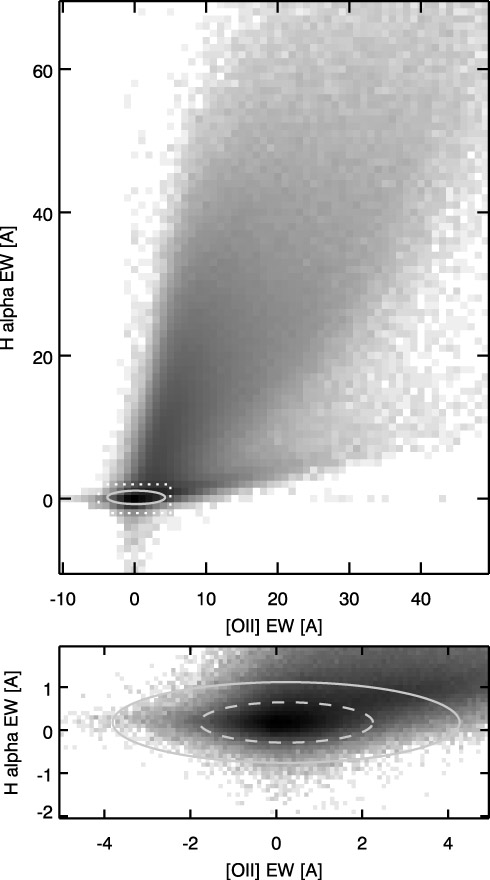}
\caption{The distribution of the 631,949 galaxies in the cross-matched VACG + MPA-JHU catalog in H$\alpha$ and [O\textsc{ii}] equivalent width. The bottom plot is a zoom in (dashed box in top plot) on our selection ellipse, shown in gray in both top and bottom figures. The bottom plot also shows the stiffer selection criteria (dashed line) that we do not use as we find them to be unnecessarily narrow.}
\label{selection}
\end{center}
\end{figure}

We include only those galaxies where emission line determinations can be made: this removes galaxies
  without measured line strengths or line strength errors, low-$S/N$
  spectra (median $S/N < 5$ \AA$^{-1}$), and those with redshift
  mismatches between the NYU VAGC reduction and the MPA-JHU reduction. This cut reduces the parent sample to 631,949 galaxies.
  We examine the equivalent width (EW) in emission of the two
  strongest optical emission lines: H$\alpha$ and the doublet
  [O\textsc{ii}]$\lambda$3727, which is unresolved in the SDSS
  spectra. Figure \ref{selection} shows the H$\alpha$-[O\textsc{ii}]
  distribution of all 631,949 galaxies in the parent sample.
  There is a peak in the distribution near zero EW in both H$\alpha$
  and [O\textsc{ii}]; these are the quiescent galaxies we wish to
  select.  There is also a narrow plume of galaxies with strong
  [O\textsc{ii}] emission but weak H$\alpha$ running almost horizontally
  rightward from the zeropoint, as well as a broader plume extending above with substantial emission in both H$\alpha$ and
  [O\textsc{ii}].  The narrow horizontal plume is dominated by
  low-ionization nuclear emission-line regions (LINERs), while the
  broad plume to the right contains star-forming galaxies, active
  galactic nuclei (AGN), and ``transition objects,'' which combine both
  star formation and AGN activity \citep{yan06}.

The lower panel of Figure \ref{selection} zooms in on the region
  around the quiescent galaxies.  We assume that the peak of the
  distribution, which lies slightly off from the origin at (0.23,0.18), represents the
  true zeropoint of the bivariate distribution, and select galaxies in
  an ellipse centered around this point (solid ellipse in Figure \ref{selection}). The ellipse has semiaxes of 4.0 \AA ~in [O\textsc{ii}] and
  0.94 \AA ~in H$\alpha$, chosen to enclose the majority of the
  quiescent objects without substantial contamination from low-level
  star-forming or LINER-like objects, with an ellipticity matching the
  contours around (0.23,0.18).  We further require that the formal errors on
  the H$\alpha$ and [O\textsc{ii}] line strengths also lie within this
  ellipse.  We experimented with using a more restrictive sample,
  represented by the dashed ellipse in the bottom panel of Figure \ref{selection}, with semiaxes half
  the size of those of the solid ellipse, and followed these data through
  the rest of the analysis.  This smaller, more restrictive sample
  produced the same results as the larger sample with poorer
  statistics.  We therefore conclude that the galaxy colors in the
  larger sample are not significantly more contaminated by galaxies with
  star-formation or AGN activity than the more selective sample and proceed with the larger sample of
  objects defined by the outer ellipse in the lower panel of Figure \ref{selection}.  Quiescent
  galaxies selected in a similar manner were used in
  \citet{graves09_paperI}; figure 1 of that work shows that these
  quiescent galaxies populate the red sequence in the color magnitude
  diagram, with only a limited number of outliers.  These criteria
  therefore select a relatively clean red sequence sample without
  applying an explicit color cut, leaving a sample of 163,101
  galaxies.

The sensitivities of the reddening corrections we measure in this work
depend strongly on the width of the underlying color distribution of
our sample galaxies.  Thus, every effort is made to identify and
exclude interlopers and color outliers, without making explicit
color cuts that will bias the intrinsic galaxy color-distribution.
Having identified a sample of quiescent, passively evolving galaxies,
we make several further cuts to keep the intrinsic color distribution
small.  These are listed and quantified in Table
\ref{selection_table}, which also lists the number of sample galaxies
remaining after each cut.  In total, the cuts remove $\sim 7$\% of the
quiescent galaxies identified in Figure \ref{selection}.  We have
confirmed that the main effect of each of these cuts is to remove
outliers from the residual color-color plot (see Figure \ref{gmrrmi}).  

The cuts include the following:  Some imaging data have substantially
larger-than-average photometry errors in the $griz$ bands, such that
they stand apart from the main locus of galaxies in a plot of the flux
errors versus the measured flux. We define cuts that exclude
these galaxies as specified in Table \ref{selection_table}, reducing
the sample size by $\sim 2$\%.  We further remove targets that have
been identified as moving objects by the SDSS pipeline by requiring
that the ``moving fit'' is good and identifies the object as
``stationary'' (another $\sim 3$\%).  Other outliers in residual
color-color space, when examined in thumbnail images, have a high
frequency of overlapping or superimposed companion objects, which
evidently affect the photometry.  We screen out such objects by
removing galaxies whose Petrosian (aperture-based) photometry differs
by $> 10$\% from the photometry based on model light profiles ($\sim
1$\% of the sample), as this turns out to be a good way of detecting
overlapping or superimposed targets.  Finally, in order to minimize the
intrinsic color spread in our standard crayon population, we remove
post-starburst galaxies (which may have substantially bluer colors) by
excluding galaxies whose spectral modeling includes more than a 20\%
flux contribution from stellar populations under 1 Gyr old ($\sim 1$\%
of the sample), based on the ``model\_coefs'' parameter in the
``gal\_indx'' table of the MPA-JHU spectral
catalog.\footnote{\url{http://www.mpa-garching.mpg.de/SDSS/DR7/SDSS\_indx.html}}
These cuts result in a final sample of 151,637 galaxies.

\begin{deluxetable}{llc}
\tabletypesize{\small}
\tablecaption{Selection Criteria\label{selection_table}}
\tablewidth{0pt}
\tablehead{
\multicolumn{2}{l}{Quiescent galaxies} &
\colhead{163,101}
}
\startdata
\multicolumn{2}{l}{Good photometry in $griz$} &159,987 \\
&$\log \sigma_g < 0.7 \log F_g - 1.2$ & \\
&$\log \sigma_r < 0.7 \log F_r - 1.2$ & \\
&$\log \sigma_i < 0.7 \log F_i - 1.15$ & \\
\vspace{0.1in} &$\log \sigma_z < 0.6 \log F_z - 0.65$ & \\
\multicolumn{2}{l}{No moving objects} &155,387 \\
&OBJC\_FLAGS2 bit 3 (BAD\_MOVING\_FIT) not set \\
\vspace{0.1in} &OBJC\_FLAGS2 bit 4 (STATIONARY) set \\
\multicolumn{2}{l}{No superimposed galaxies} &153,676 \\
\vspace{0.1in} &$PetroFlux_r/ModelFlux_r < 1.1$ & \\
\multicolumn{2}{l}{No post-starburst galaxies} &151,637 \\
\vspace{0.1in} &$f_{young} < 20$\% & \\
\hline \\
\multicolumn{2}{l}{Final clean quiescent galaxy sample} &151,637 \\
\enddata
\end{deluxetable}

\section{Standard Crayons}\label{sc}

The sample described above provides a set of galaxies
  that form a well-defined red sequence and can therefore be used
  as standard crayons to detect reddening due to foreground dust in
  the Milky Way.  There are, however, several remaining sources of
  color variation that could produce structure in the color residual
  map, either through direct spatial correlations, or through
  correlations with LSS or redshift.  We must
  identify and account for these variations as well as possible, then
  test the final data map for lingering spurious variations.

These variations come in two cateogries: (1)
  spatially-dependent errors or selection biases and (2) intrinsic
  color variations within the galaxy population that correlate with
  other galaxy properties.  
    Other sources of error are more difficult to correct, but
  we can minimize their impact in the final residual map.  We discuss
  these in \S \ref{lss_errors} and \ref{phot_zeropoint}, where we assess
  various sources of error in the residual maps.

In this section, we consider intrinsic color
  variations due to several effects.  These include the
  color-magnitude relation, environment-driven variations in galaxy
  colors at fixed mass, redshift-dependent sources of
  color variation, and photometric errors.  We then describe our method for fitting out these
  variations so that they do not bias the final residual map.

\subsection{The Color-Magnitude Correction}\label{cmc}

The most obvious example of intrinsic color variation
  is the color-magnitude relation (CMR), in which more luminous
  galaxies have redder colors.  There is also clear evidence that the most
  luminous red sequence galaxies are found in higher-density
  environments \citep{hogg03}.\footnote{In fact, the distribution of
    typical environments along the red sequence is complicated,
    showing a ``saddle'' shape, with $\sim L^{\star}$ galaxies
    typically found in less dense environments than either more or
    less luminous red sequence galaxies.  The existence of LSS
  therefore couples the known color variations as a function of
  luminosity to the distribution of galaxies on the sky.  A massive
  cluster, with a concentration of very luminous, very red galaxies
  would produce the effect of a highly reddened sight-line if this
  effect is not accounted for.} Additionally there is a subtle
variation of the CMR with redshift; galaxies at the extremes of our
redshift range have colors more strongly correlated with intrinsic
brightness. This variability in the CMR is likely
influenced by the brightness limitations (Malmquist bias) of the survey.
This shifting relationship may not be due to genuine intrinsic variations in
  the galaxy properties, but it nonetheless must be taken into account when
  constructing our reddening maps. We note that as long as the variations observed in the surveyed galaxies are taken into account, the difference between genuine galaxy color evolution and the effects of observation bias will not affect our final color residuals. 

\subsection{The Color-Environment Correction}\label{cec}

There is a well-known relationship between galaxy color and various
measures of density of galactic environment. There is a similar
relationship within our subsample of passive galaxies, with a slight
trend of galaxies in dense environments being redder. To infer
environmental density of a galaxy, we use the $\delta_5$ parameter of
\citet{cooper08} for each galaxy. The $\delta_5$ parameter is defined as the surface density of
spectroscopically selected SDSS galaxies within 1000 km s$^{-1}$ and
angularly closer than the 5th nearest neighboring galaxy, normalized
by redshift. We find that a simple first-order polynomial fit captures
all the variation of color with $\delta_5$.

\subsection{Method of Correction}\label{mc}
As will be justified in \S \ref{pos}, we fit for the variation of color with magnitude and environment using a median (minimum absolute residual) fit. Specifically we fit the data with
\beq\label{fit}
a-b = c \alpha\left(z\right) + \delta_5 \beta\left(z\right) + \gamma\left(z\right)
\eeq
where $a-b$ represents an arbitrary K-corrected, reddening corrected color, $c$ an arbitrary K-corrected, reddening-corrected absolute magnitude, and $\alpha$, $\beta$ \& $\gamma$ are 8th-order polynomial functions of redshift, $z$, e.g. :
\beq
\alpha\left(z\right) = \sum_{i=0}^8f_{i}z^i.
\eeq
We are minimizing the sum of the absolute residuals in $a-b$, $c$, and $\delta_5$, with the assumption of 1\%, uncorrelated errors in each. This 27-parameter fit is executed with MPFIT (\citei{Markwardt09}; an implementation of the Levenberg-Marquardt minimization), first in the standard implementation, to find a ``least-squares'' fit, and subsequently in a modified implementation to find a minimum absolute residual fit. 

\subsection{Photometric corrections}\label{pc}
The SDSS data are marvelously uniform, with excellent photometric corrections, but, at the level of our investigation, the remaining errors in the photometry can be important. Once the intrinsic color relationships described in \S  \ref{cmc} and \S \ref{cec} have been removed from the galaxy data, our sample allows us to investigate photometric errors in SDSS that are uncorrelated to the expected structure of fluctuations in the SFD98 map. 

The SDSS camera \citep{Gunn98} contains 6 columns of CCDs, each with 5
CCDs for the 5 SDSS filters. Each CCD covers an area of sky
13$^\prime$ wide as the camera drifts across the sky in a series
of strips, each dozens of degrees long. Since the filter properties vary slightly from column to column \citep{Doi10}, we compare the median of our observed galaxy colors as a function of camera column, under the assumption that there is no intrinsic structure in the errors in the SFD98 reddening map in this particular pattern on the sky. We indeed find that our galaxy colors do vary detectably by column from the median color, with a range of $-0.016$ magnitudes ($u-z$, column 2) to $+0.027$ magnitudes ($u-z$, column 5) and a typical amplitude of 0.008 magnitudes. We remove these spurious photometric variations from our galaxy sample.

\subsection{Properties of the Sample}\label{pos}
\begin{figure}
\begin{center}
\includegraphics[scale=1., angle=0]{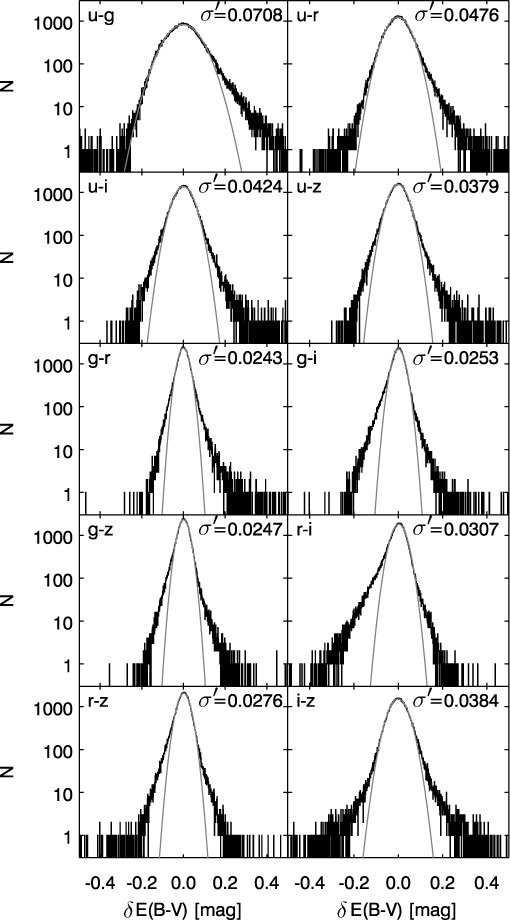}
\caption{Histograms of the error in selective extinction as determined from colors of each galaxy, shown on a log scale. One histogram is shown for each of the 10 SDSS colors. The Gaussian fit is overplotted, and the width of that Gaussian noted in the upper right of the plot. The non-Gaussianity typically begins at about 10\% of the peak of the distribution.}
\label{sigprime}
\end{center}
\end{figure}

\begin{figure}
\begin{center}
\includegraphics[scale=0.4, angle=0]{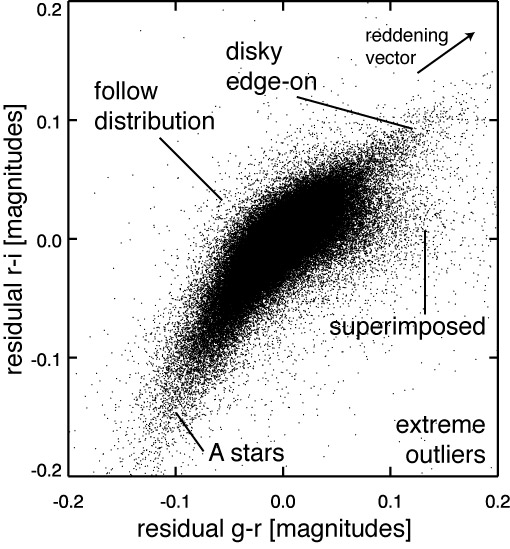}
\caption{The residual color distribution of the galaxies, in $g-r$ vs. $r-i$ colors. We identify 5 contributions to non-Gaussianity in the distribution: internal reddening in disky, edge-on galaxies; superpositions of quiescent galaxies; extreme outliers; galaxies that contain significant A stars; and a component that follows the main distribution. See \S \ref{pos} for details.}
\label{gmrrmi}
\end{center}
\end{figure}

After applying the corrections, we are left with a distribution of residual galaxy colors. Since in principle, any of the 10 color pairs that can be formed from the 5 SDSS bands can be used to determine the errors in the SFD98 reddening map, we are interested in the properties of each of these residual colors. Firstly, we note that none of the colors have distributions that are Gaussian (see Figure \ref{sigprime}). All distributions have wide wings that dominate the standard deviation, although Gaussian fits are relatively accurate for most colors down to about 10\% of the peak of the histogram. We identify five separate sources of this non-Gaussianity, although we do not attempt to fully characterize them. Each of these sources can be seen in the residual $g-r$ vs.~ residual $r-i$ scatter plot, Figure \ref{gmrrmi}. 

The first comes from interloper galaxies with residual populations of A stars, which form the ``plume'' towards the bottom left of the plot. These are galaxies that had a recent burst of star formation, but are no longer forming stars. Note that this plume is dramatically reduced, but not removed, by the post-starburst cut implemented in \S \ref{selection_criteria}. The second comes from galaxies with disky components that are viewed mostly edge on, and thus are significantly obscured by their own dust; this group lies directly along the reddening vector. The third comes from errors in the determination of the galaxy flux. If the galaxy is superimposed with another galaxy of similar type the color will be unaffected but brightness will be increased. This error will lead to an incorrect color-magnitude correction, which will affect the residual color. This is manifest as a plume toward positive $g-r$. This effect is much stronger in $g-r$ than $r-i$, as the CMR has a much stronger dependence in $g-r$ and thus the error in the correction is larger. Note that this plume is dramatically reduced by the photometric fit cuts in \S \ref{selection_criteria}. The fourth source is a population of extreme outliers, with no particular position in color-color space. The fifth group of objects that do not follow a Gaussian color distribution, and which dominate by number, seem to simply follow the standard distribution of the data. It is not clear why the main color distribution has these ``natural'' non-Gaussian wings, though they may stem from non-Gaussian errors in the photometry itself.

In light of these sources of non-Gaussianity, we explore the use of
median (least-absolute-residual) fitting. Median fitting has the
inherent disadvantage of typically having less overall accuracy; in a
Gaussian distribution the error in the median is $\sqrt{\pi/2} \simeq
1.253$ times larger than the error in the
mean. However, bootstrap resampling of the residual
  galaxy colors shows that the statistical error in the median has a
  $\sim 5\%$ \emph{smaller} scatter than in the mean due to the
  non-Gaussian distribution of galaxy color residuals. We could use an outlier-rejecting method, such as
$\sigma$-clipping, to reduce the error in the mean. Since the color
distributions seem to have intrinsically non-Gaussian shapes, and it
is difficult to determine the difference between outliers and members
of the distribution, we choose to rely on median fits throughout this
work.

Despite the non-Gaussian wings of the color residual
  distributions, the bulk of the galaxies do follow a Gaussian
  distribution, and it is these galaxies that dominate the uncertainty
  in the median.  We therefore fit the color residual histogram with a
  Gaussian plus an overall vertical offset (to account for the extreme
  outliers) and use the standard deviation of the core Gaussian
  profile to calculate the uncertainty in the median. We denote the fitted standard deviation as
$\sigma^\prime$.  

In theory, our final reddening correction map can be
  made based on any of the measured residual colors; in practice it
  makes sense to choose the color residual with the narrowest
  distribution and therefore the lowest uncertainty in the median.
To compare the distributions among colors, we use the figure of merit
$\delta E \left(B-V\right)$, the error in the selective extinction
determined from the color of a galaxy. For a given color, $b-c$,
corrected by an arbitrary magnitude band, $a$, we find
\beq\label{delE} \delta
E\left(B-V\right)_{b-c,a}=\frac{\delta\left(b-c\right)}{f\left(b\right)-f\left(c\right)-f\left(a\right)
  \alpha\left(z\right)}, \eeq where $\delta\left(b-c\right)$ is the
residual color found after applying the corrections is \S \ref{mc},
$f\left(a\right)$ is the ratio of extinction in band $a$ to
$E\left(B-V\right)$, and $\alpha\left(z\right)$ is the slope of the
color-magnitude relationship, as in Equation \ref{fit}. Note that this
relationship is somewhat complicated due to the non-trivial projection
of the reddening vector onto the color-magnitude relationship.  Figure
\ref{sigprime} shows the distribution of all ten $\delta
E\left(B-V\right)_{b-c,r}$. The equivalent distributions using other bands as the reference magnitude (i.e.,
  $\delta E\left(B-V\right)_{b-c,g}$, $\delta
  E\left(B-V\right)_{b-c,i}$ and $\delta E\left(B-V\right)_{b-c,z}$,
  not shown) are nearly identical, though marginally broader. We find
that all $\delta E \left(B-V\right)$ distributions using the $u$
  band (i.e., $\delta E\left(B-V\right)_{u-c,r}$) are relatively
broad due to two effects: (1) the larger photometric
  errors in the SDSS $u$ band (see e.g., \"ubercal, \citealt{Doi10}), and (2) the
  larger intrinsic scatter in $u$ due to to variations in the stellar
  populations (because the $u$ band is the most sensitive to small fractions
  of recent star formation).  Colors involving the $u$ band are therefore not suitable for our purposes.  All else being equal, we would use $g-r$ to determine reddening, as it has a marginally smaller fitted standard deviation in $\delta E \left(B-V\right)$ than other colors. We note that combining these colors does not produce a reduced $\sigma^\prime$.

\section{Extinction Correction Maps}\label{ecm}
We wish to produce a map that covers the SDSS DR7 contiguous spectroscopic area towards Galactic zenith using the residual galaxy colors we generated in \S \ref{sc}. 139,526 of the 151,637 galaxies lie in this part of the SDSS sky.

\subsection{Method of Map Construction}\label{mmc}

Since we do not have a galaxy for every position in the sky
for which we wish to compute a reddening correction, we must apply
some amount of smoothing. This situation is analogous to a problem that
confronts single-dish radio telescopes, wherein a large number of telescope pointings are made, recovering
information only near that single direction. These time-ordered data
(TOD) are generally taken in a non-homogenous pattern in the sky, and
they must be combined to form a usable grid. In our case the galaxies
play the role of the TOD, and we attempt to construct a smooth grid of
reddening correction. Typically, for any point in the grid, the values
of nearby TODs are weighted more heavily, and distant TODs more
weakly, with some kind of effective point spread function (PSF). In
this work we use a Gaussian PSF. In contrast to the problem confronting radio telescopes, we are only collecting information from the angular area toward the galaxy itself, rather than a large radius around the direction in which the antenna is pointed, typically larger than the spacing between pointings. This undersampling of the sky limits the fidelity of the maps we construct (see \S \ref{souerr}). 

Since we wish to use medians, rather than means, to determine the
reddening error, we have modified the single-dish code used in
\citet{PH08}, SDGRID, to allow for weighted median fits. The weighted
median, $\tilde{x}$, is the value that minimizes \beq
\frac{\sum_{i=1}^N|x_i-\tilde{x}|w_i}{\sum_{i=1}^N w_i}, \eeq Where
$x_i$ is the set of values at each point (i.e.,
$\delta E\left(B-V\right)_{b-c,a}$ at each galaxy) and $w_i$ is the
set of weights, in this case determined by the PSF. We have also
modified SDGRID to determine the error in the weighted median at each
pixel.  The value of each map pixel is determined by a different $w_i$
from the particular spatial distribution of galaxies in the vicinity. $w_i$
varies due to random spatial fluctuations, edge effects, and the
clustering of galaxies in LSS. To compute the error in the weighted
median for each point we use a bootstrap method. First, we assume that the distribution of values contributing to each point, $x_i$, is relatively similar. This assumption allows us to replicate the measurement of $\tilde{x}$ with a large number
of sets of values $x_i$ randomly selected from the data set, but using
the same $w_i$. We are then able to determine the typical error in
$\tilde{x}$. We note that our assumption of identical distributions is
certainly incorrect (were it strictly true we would see no signal in
our map) but that since the random galaxy-to-galaxy scatter
($\sigma^\prime \simeq$ 0.03 magnitudes $E(B-V)$) dwarfs the map
signal ($\sigma^\prime \simeq$ 0.005 magnitudes $E(B-V)$), it is a
reasonable and conservative approximation, as it only slightly overestimates our errors.

Since we are constructing a map that covers a large area of the sky, we must take care in the choice of mapping projection. Following SFD98, we use the Zenithal Equal Area (ZEA) projection, with a projection center at $b=90^\circ$. Since our data are all at high Galactic latitude, being projected from zenith seems a minimally distorting choice, and since the pixels in ZEA projections have equal area, we need not be concerned that data in different pixels will have systematically different error properties. 

In addition to choosing a PSF for our mapping routine, we must also choose a resolution. There are two competing forces in our selection of resolution. At high resolution, we are able to capture small scale fluctuations in the reddening error, thus increasing the fidelity of the measurement. At lower resolution we have more galaxies contributing to a given measurement, thus increasing S/N. If we assume that all spatially correlated variations in galaxy colors stem from the errors in Galactic reddening, we can find an optimal scale with the following Monte Carlo test. 

We use 90\% of the galaxies to build a map at some resolution, $\Theta$, where $\Theta$ is the full-width at half-maximum (FWHM) of the Gaussian PSF. We then take the remaining 10\% of the galaxies and determine the extent to which applying the generated map reduces the scatter:
\beq
\xi_\Theta = \frac{\sigma^\prime\left(\delta E\left(B-V\right)_{b-c,a}-\langle{\delta E\left(B-V\right)_{b-c,a}}\rangle_\Theta \right)}{\sigma^\prime\left(\delta E\left(B-V\right)_{b-c,a}\right)}
\eeq
Here $\langle x \rangle_\Theta$ is a map generated using the value $x$ with a PSF with FWHM = $\Theta$. To produce reliable values we repeat the experiment 200 times and use the aggregate data points to compute $\xi_\Theta$. We find that, as expected, $\xi_\Theta$ increases toward small and large values of $\Theta$, with a broad minimum between $\Theta \simeq 3^\circ$ and $\Theta \simeq 5^\circ$. 

\subsection{Sources of Error in the Map}\label{souerr}
The beauty of our particular set of standard crayons is that while there could easily be complex variation in the colors of the galaxies, based on their metallicity, star formation history (see, e.\ g.\ , Figure \ref{gmrrmi}), nuclear activity, and internal reddening, none of these depend upon their position on the sky. Thus, we can ignore these variations and treat them simply as contributions to the random ``cosmic'' scatter, which should degrade, but not skew, our results. There are, however, a few concerns that can skew our data, where systematic errors can have an effect on the spatial structure on the sky.

\subsubsection{Coupling to Large Scale Structure}\label{lss_errors}
In a strange role reversal, in this work the LSS manifests as an annoying systematic that gets in the way of our study of reddening, rather than the other way around. We have already dealt with one effect of LSS: the color-density relation discussed \S \ref{cec}. 

A second effect of clustering is that it couples sky positions to specific redshifts. This correlation can be problematic if there is variation in the color of galaxies with small changes in redshift, as it can contribute to erroneous fluctuations in the map. We know of two possible causes of variations in color as a function of very small changes in redshift ($\Delta z \sim 0.001$): K-correction effects and skyline effects. 
 
 It is possible that the applied K-correction may be systematically under- or over-correcting for the effect of the galaxy spectrum moving from filter to
filter as a function of redshift, 
contaminating our sample. Skyline effects can also
impact our galaxy colors as a function of small changes in
redshift. The sky shines brightly with various molecular lines,
particularly hydroxyl. At redshifts where either of our two
selection-criteria lines ([O {\sc
    II}] or H$\alpha$) align with the skylines, the
errors in our determination of those lines will dramatically
increase. The primary effect of skyline alignment is to remove
galaxies from our sample that we would otherwise select, because they
do not meet the limits on EW errors in H$\alpha$. Strong skylines may also allow
star-forming galaxies to leak into the sample, if the sky is
over-subtracted.  To determine the effect of these fluctuations on our
final map, we re-assign each galaxy the color expected from the median color in its $\Delta z =
0.001$ redshift bin. We then rebuild the map with these redshift
determined colors. We find that this map does have structure, but the
variation is very small compared to the error in our true reddening
map: the median absolute variations are only 0.07$\sigma$, with no
points larger than 0.6$\sigma$. We therefore can ignore the impact of
fluctuations in color with small shifts in redshift.
 
The third effect from LSS we consider is additional unmodeled noise in the map. To illustrate this effect, we propose the following situation. Galaxies that contribute weight to map pixel A are more clustered on the sky than those that contribute to map pixel B. Imagine the true sky reddening has significant  fluctuation from the SFD98 reddening map below the resolution of our map. This sub-resolution fluctuation will add error to both pixels, but since the errors are significantly more correlated from galaxy to galaxy in pixel A, the error will be higher in pixel A. To test the effects of this, we make a map, using the method of \S \ref{mmc}, of the angular distance to the nearest neighboring galaxy $\phi_{min}$. We then examine the reduced $\chi^2$ of our reddening map as a function of this ``clustering map'', $\langle \phi_{min} \rangle_\Theta$. We find that, even for small values of $\Theta$ where we expect this effect to be strongest, there is no evidence for increased noise with increased clustering.

\subsubsection{Large-scale striation from systematic photometric errors}\label{phot_zeropoint}
\begin{figure*}
\begin{center}
\includegraphics[scale=0.8, angle=0]{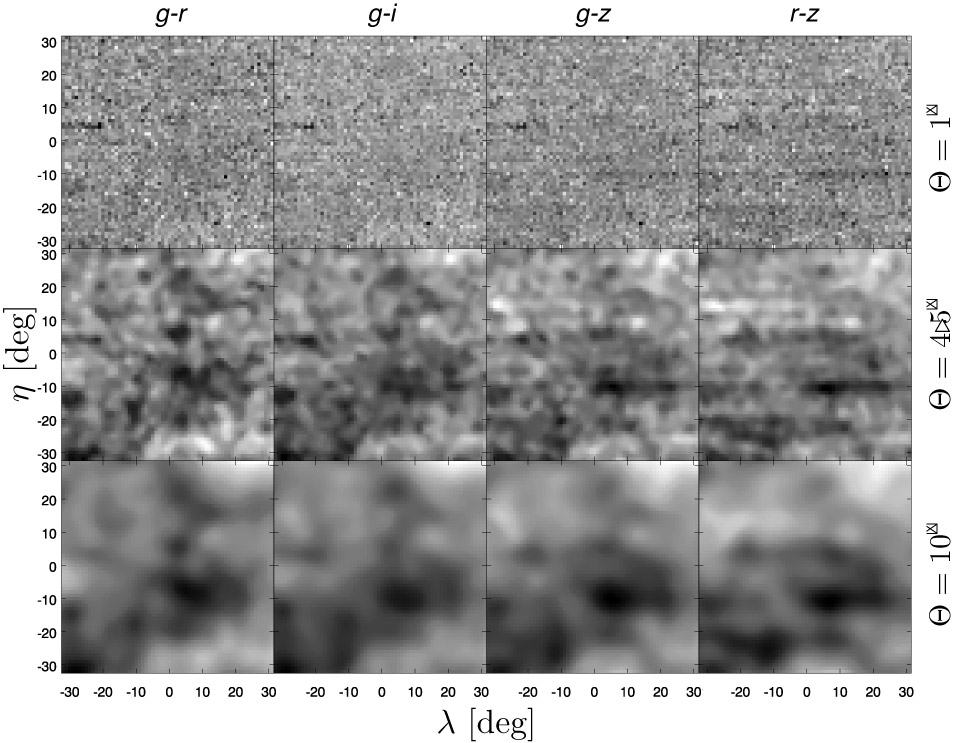}
\caption{Map of the redding in SDSS coordinates $\eta$ and $\lambda$ over the largest contiguous square area of the SDSS Northern Galactic cap. We show reddening as predicted from the 4 color pairs that have $\sigma^\prime < 0.03$ and for 3 different values of $\Theta$. Note the striation in redder color pairs and for lower values of $\Theta$.}
\label{stripe}
\end{center}
\end{figure*}

\begin{figure*}
\begin{center}
\includegraphics[scale=0.8, angle=0]{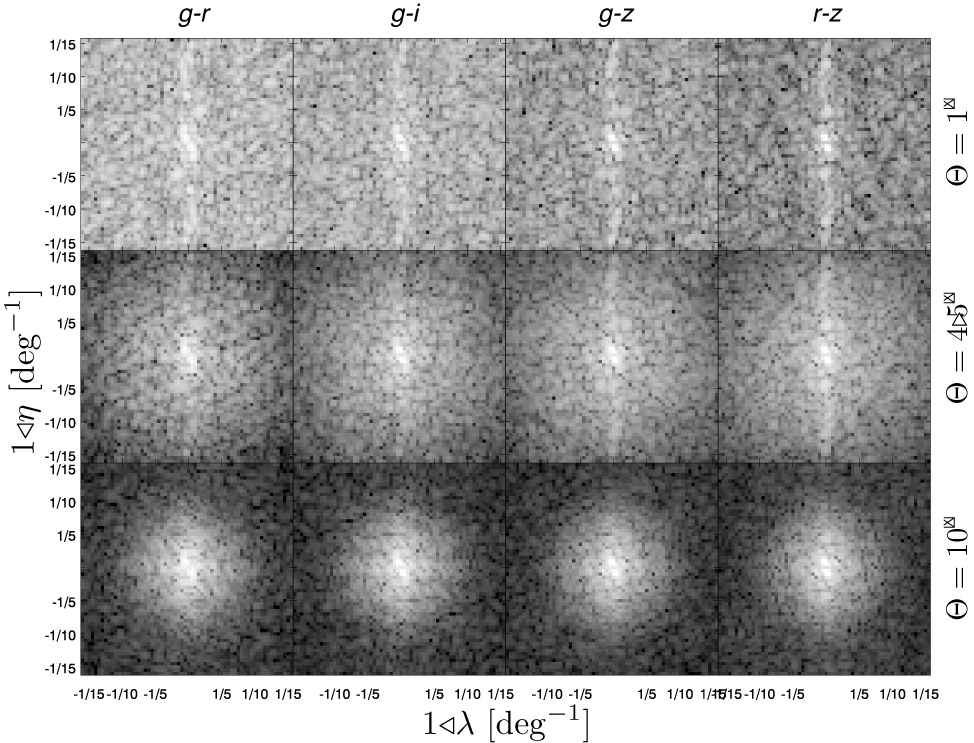}
\caption{Same as Figure \ref{stripe}, but transformed to the Fourier domain. Again, note the prominent striping at high wavenumber for redder color pairs and lower values of $\Theta$.}
\label{stripeft}
\end{center}
\end{figure*}

 \"Ubercal made great strides in reducing and characterizing the photometric errors in the SDSS data. While \"ubercal reduced the errors to less than 1\% in $g$,$r$,$i$, and $z$ bands, it also uncovered very small systematic variations that depended upon the stripes in the data that it could not correct. SDSS photometry data are taken in long tracks, and \"ubercal found that while it did significantly reduce the variation in photometric values from track to track, detectable striations remained (see \S 5.4 and figure 13 of \"ubercal), especially in $z$ band. To examine the effect of these striations in our data, we make maps at a variety of resolutions in SDSS $\lambda$ and $\eta$ coordinates, where $\lambda$ is along the scan direction and $\eta$ is across the scan direction (see Figure \ref{stripe}). We then take the largest fully sampled square of data, and examine the Fourier power in 2D, using the method described in SFD98, \S 3.1.1 to avoid edge artifacts (Figure \ref{stripeft}). Consistent with the findings of \"ubercal we find that the only detectable striations occur along the $\lambda$ axis and that colors involving $z$ are significantly more striated that those involving only $g$, $r$, and $i$. In particular we find that all maps have detectable striation at $\Theta = 1^\circ$, no maps have detectable striation at $\Theta = 10^\circ$, and only the $g-r$ maps lack striation within the optimum range $3^{\circ} < \Theta < 5^{\circ}$ found in \S \ref{mmc}. We therefore adopt $\Theta = 4.5^\circ$ as our map resolution and $g-r$ as the color by which we measure $E(B-V)$. 

\section{Structure of the Extinction Correction Maps}\label{results}
Our final dust correction maps (and associated errors) are shown in Figure \ref{map}, and are made available to the public at \url{http://www.peekandgraves2010.com}. We detect variations from the SFD98 reddening map at greater than 3$\sigma$ over large areas of the map, with detection greater than 10$\sigma$ in regions toward the Galactic anti-center at lower latitudes. The map has 1$\sigma$ sensitivity of 1.5 millimagnitudes in $E\left( B-V\right)$, with maximum deviation of 45 millimagnitudes toward $\alpha = 128^\circ$, $\delta = +62^\circ$. We note that this position is coincident with both the ultra-faint dwarf Ursa Major II and a bright knot in the High-Velocity Cloud (HVC) complex A. While it is possible that this galaxy or HVC complex is contributing to error in the Galactic reddening, we find no association between our residual reddening map and either other ultra-faint dwarfs or other HVC complexes. We therefore expect that this association is simply a coincidence. 

While deviations from SFD98 are small in absolute scale, they can be fractionally large, exceeding 50\% in excess or decrement in some areas. On the whole, the map also shows that towards lower latitudes (and higher extinction) the SFD98 map underpredicts extinction relative to higher latitudes. This trend toward underprediction of reddening in regions of increased dust is not in disagreement with the claims that SFD98 overpredicts the dust in high extinction regions (e.g. \citei{AG99}, \citei{Stanek99}, \citei{Cambresy01}, \citei{YFS07}, and \citei{RF09}): we do not have data in regions with extinction above $E\left(B-V\right) = 0.15$, and these works typically examine areas with $E\left(B-V\right) \gtrsim 0.5$. We do not find obvious correlations of the map with the IRAS 100 $\mu$m map, Galactic \hi column density maps \citep{Kalberla2005}, or the LSS in the SDSS.


It is also worth noting that the application of our correction to the SFD98 maps results in only 40 square arcminutes ($\sim$7 SFD98 map pixels) of area with a negative reddening correction, the largest of which is only $-0.3$ millimagnitudes $E\left( B-V\right)$, far smaller than our error bars. For these very few areas we report a final reddening correction of zero. This result can be taken as at least a partial validation of the SFD98 zero-point; were the zero-point set incorrectly too low we would expect to find large areas with negative reddening.

\subsection{Correlation With Dust Temperature Map}
\begin{figure*}
\begin{center}
\includegraphics[scale=0.5, angle=0]{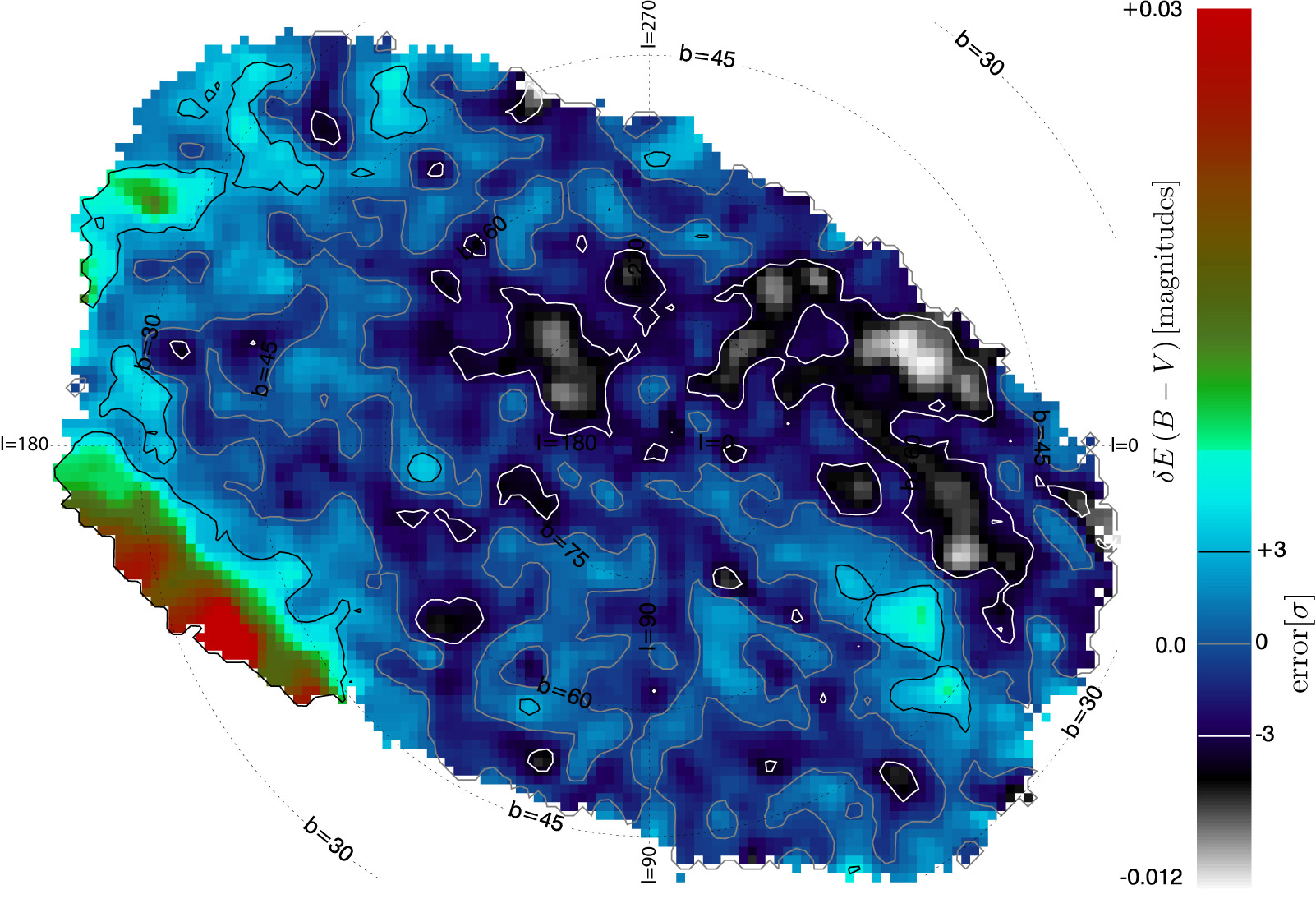}
\caption{Our final extinction correction map. We show both our extinction correction and our significance contours. Galactic longitude and latitude are overlaid, with Galactic zenith at the middle of the map and Galactic center off the right side of the map. Note that the sign of this map is such that it portrays the residual errors in the SFD98 map, or that which must be subtracted to improve the SFD98 map.}
\label{map}
\end{center}
\end{figure*}

\begin{figure*}
\begin{center}
\includegraphics[scale=0.5, angle=0]{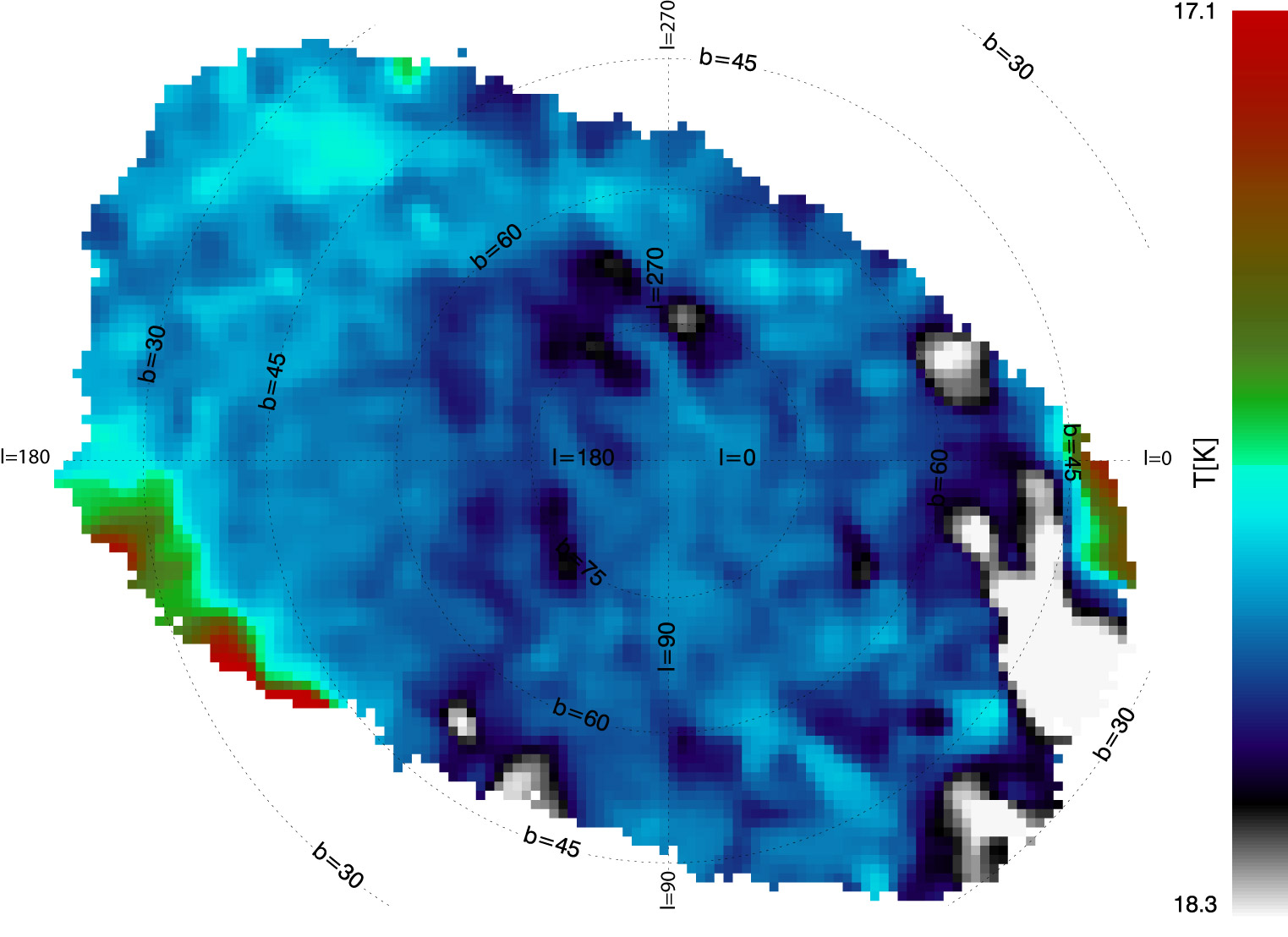}
\caption{Temperature map of SDSS DR7 contiguous area, generated by finding the temperature value in the DIRBE map toward each of our galaxies and then applying the same method as used for our reddening correction map to facilitate comparison. Red represents \emph{cooler} areas. Note the similarity to Figure \ref{map}, in regions away from Galactic center (left side of the map). }
\label{tmap}
\end{center}
\end{figure*}

For comparison to our reddening correction maps, we generate a map of the dust temperature. This map is generated by determining the dust temperature at each galaxy position using the SFD98 temperature map, as derived from the DIRBE instrument data. We then take these values and apply the identical map construction algorithm described in \S \ref{mmc}. This map is shown in Figure \ref{tmap}. We note that this map, with red representing \emph{cooler} dust, is very similar map shown in Figure \ref{map}, with features reproduced down to the resolution scale. This correlation implies that the SFD98 map underpredicts reddening in regions of low dust temperature. In the context of Equation \ref{expo}, $X\left(R \right)$ is too small for regions of relatively low $R$. We note that this similarity does not persist throughout the entire map, diverging most prominently toward Galactic center. Perhaps this divergence indicates that the $X\left(R \right)$ function used is more appropriate for that area of sky than toward other areas. It is also possible that the superposition of grains of different temperatures (see figure 2 of SFD98) or varying grain morphology or composition could be the culprit. The correlation between our residual map and the DIRBE temperature map is also strong evidence for the claim that we are indeed measuring errors in the reddening map, rather than an unrelated systematic.

\section{Conclusions}\label{conc}
In this work we used quiescent galaxies as ``standard crayons'', and found errors in the SFD98 reddening map at high Galactic latitude. We found that a pure spectroscopic selection of quiescent galaxies with no H$\alpha$ or [O\textsc{ii}] emission produced a very homogeneous population of galaxies. Once these galaxies were corrected for redshift-dependent color-magnitude relations and  color-density relations, we found they had a scatter of less than 3\% in the bulk of galaxy colors. 
We determined that since there were significant departures from an intrinsic Gaussian color scatter, a median fitting method was desirable over a standard fit. We found that systematic errors that stem from large-scale structure in the universe did not impact our data, and that our data were relatively robust to striation from systematic errors in the SDSS photometry at a resolution of $4.5^\circ$ and using $g-r$ colors. Our final correction maps showed that SFD98 maps have errors ranging from $45$ to $-10$ millimagnitudes in $E(B-V)$, although the bulk of the map has errors less than 3 millimagnitudes. These maps are provided to the public at \url{http://www.peekandgraves2010.com}. We find the largest errors in the SFD98 map are towards lower latitudes. We find some noticeable correlation with the temperature map SFD98 derived from DIRBE data, particularly toward the outer galaxy, suggesting that the function relating temperature to emission correction factor employed in SFD98 was in error for dust grains toward the outer Galaxy at high latitudes.

\acknowledgements

Funding for the SDSS and SDSS-II has been provided by
  the Alfred P. Sloan Foundation,
  National Science Foundation, the U.S. Department of Energy, the
  National Aeronautics and Space Administration, the Japanese
  Monbukagakusho, the Max Planck Society, and the Higher Education
  Funding Council for England. The SDSS Web Site is
  \url{http://www.sdss.org/}.

The SDSS is managed by the Astrophysical Research
  Consortium for the Participating Institutions. The Participating
  Institutions are the American Museum of Natural History,
  Astrophysical Institute Potsdam, University of Basel, University of
  Cambridge, Case Western Reserve
  Drexel University, Fermilab, the Institute for Advanced Study, the
  Japan Participation Group, Johns Hopkins University, the Joint
  Institute for Nuclear Astrophysics, the Kavli Institute for Particle
  Astrophysics and Cosmology, the
  Academy of Sciences (LAMOST), Los Alamos National Laboratory, the
  Max-Planck-Institute for Astronomy (MPIA), the Max-Planck-Institute
  for Astrophysics (MPA), New Mexico State University, Ohio State
  University, University of Pittsburgh, University of Portsmouth,
  Princeton University, the United States Naval Observatory, and the
  University of Washington.  
  
JEGP wishes to thank Doug Finkbeiner and Eddie Schlafly for very hepful conversations regarding both SDSS and SFD98 and for their hospitality. 

\bibliographystyle{apj}

\begin{thebibliography}{29}
\expandafter\ifx\csname natexlab\endcsname\relax\def\natexlab#1{#1}\fi

\bibitem[{{Adelman-McCarthy} {et~al.}(2008){Adelman-McCarthy}, {Ag{\"u}eros},
  {Allam}, {Allende Prieto}, {Anderson}, {Anderson}, {Annis}, {Bahcall},
  {Bailer-Jones}, {Baldry}, {Barentine}, {Bassett}, {Becker}, {Beers}, {Bell},
  {Berlind}, {Bernardi}, {Blanton}, {Bochanski}, {Boroski}, {Brinchmann},
  {Brinkmann}, {Brunner}, {Budav{\'a}ri}, {Carliles}, {Carr}, {Castander},
  {Cinabro}, {Cool}, {Covey}, {Csabai}, {Cunha}, {Davenport}, {Dilday}, {Doi},
  {Eisenstein}, {Evans}, {Fan}, {Finkbeiner}, {Friedman}, {Frieman},
  {Fukugita}, {G{\"a}nsicke}, {Gates}, {Gillespie}, {Glazebrook}, {Gray},
  {Grebel}, {Gunn}, {Gurbani}, {Hall}, {Harding}, {Harvanek}, {Hawley},
  {Hayes}, {Heckman}, {Hendry}, {Hindsley}, {Hirata}, {Hogan}, {Hogg}, {Hyde},
  {Ichikawa}, {Ivezi{\'c}}, {Jester}, {Johnson}, {Jorgensen}, {Juri{\'c}},
  {Kent}, {Kessler}, {Kleinman}, {Knapp}, {Kron}, {Krzesinski}, {Kuropatkin},
  {Lamb}, {Lampeitl}, {Lebedeva}, {Lee}, {Leger}, {L{\'e}pine}, {Lima}, {Lin},
  {Long}, {Loomis}, {Loveday}, {Lupton}, {Malanushenko}, {Malanushenko},
  {Mandelbaum}, {Margon}, {Marriner}, {Mart{\'{\i}}nez-Delgado}, {Matsubara},
  {McGehee}, {McKay}, {Meiksin}, {Morrison}, {Munn}, {Nakajima}, {Neilsen},
  {Newberg}, {Nichol}, {Nicinski}, {Nieto-Santisteban}, {Nitta}, {Okamura},
  {Owen}, {Oyaizu}, {Padmanabhan}, {Pan}, {Park}, {Peoples}, {Pier}, {Pope},
  {Purger}, {Raddick}, {Re Fiorentin}, {Richards}, {Richmond}, {Riess}, {Rix},
  {Rockosi}, {Sako}, {Schlegel}, {Schneider}, {Schreiber}, {Schwope}, {Seljak},
  {Sesar}, {Sheldon}, {Shimasaku}, {Sivarani}, {Smith}, {Snedden}, {Steinmetz},
  {Strauss}, {SubbaRao}, {Suto}, {Szalay}, {Szapudi}, {Szkody}, {Tegmark},
  {Thakar}, {Tremonti}, {Tucker}, {Uomoto}, {Vanden Berk}, {Vandenberg},
  {Vidrih}, {Vogeley}, {Voges}, {Vogt}, {Wadadekar}, {Weinberg}, {West},
  {White}, {Wilhite}, {Yanny}, {Yocum}, {York}, {Zehavi}, \&
  {Zucker}}]{adelman-mccarthy08}
{Adelman-McCarthy}, J.~K., {Ag{\"u}eros}, M.~A., {Allam}, S.~S., {Allende
  Prieto}, C., {Anderson}, K.~S.~J., {Anderson}, S.~F., {Annis}, J., {Bahcall},
  N.~A., {Bailer-Jones}, C.~A.~L., {Baldry}, I.~K., {Barentine}, J.~C.,
  {Bassett}, B.~A., {Becker}, A.~C., {Beers}, T.~C., {Bell}, E.~F., {Berlind},
  A.~A., {Bernardi}, M., {Blanton}, M.~R., {Bochanski}, J.~J., {Boroski},
  W.~N., {Brinchmann}, J., {Brinkmann}, J., {Brunner}, R.~J., {Budav{\'a}ri},
  T., {Carliles}, S., {Carr}, M.~A., {Castander}, F.~J., {Cinabro}, D., {Cool},
  R.~J., {Covey}, K.~R., {Csabai}, I., {Cunha}, C.~E., {Davenport}, J.~R.~A.,
  {Dilday}, B., {Doi}, M., {Eisenstein}, D.~J., {Evans}, M.~L., {Fan}, X.,
  {Finkbeiner}, D.~P., {Friedman}, S.~D., {Frieman}, J.~A., {Fukugita}, M.,
  {G{\"a}nsicke}, B.~T., {Gates}, E., {Gillespie}, B., {Glazebrook}, K.,
  {Gray}, J., {Grebel}, E.~K., {Gunn}, J.~E., {Gurbani}, V.~K., {Hall}, P.~B.,
  {Harding}, P., {Harvanek}, M., {Hawley}, S.~L., {Hayes}, J., {Heckman},
  T.~M., {Hendry}, J.~S., {Hindsley}, R.~B., {Hirata}, C.~M., {Hogan}, C.~J.,
  {Hogg}, D.~W., {Hyde}, J.~B., {Ichikawa}, S.-i., {Ivezi{\'c}}, {\v Z}.,
  {Jester}, S., {Johnson}, J.~A., {Jorgensen}, A.~M., {Juri{\'c}}, M., {Kent},
  S.~M., {Kessler}, R., {Kleinman}, S.~J., {Knapp}, G.~R., {Kron}, R.~G.,
  {Krzesinski}, J., {Kuropatkin}, N., {Lamb}, D.~Q., {Lampeitl}, H.,
  {Lebedeva}, S., {Lee}, Y.~S., {Leger}, R.~F., {L{\'e}pine}, S., {Lima}, M.,
  {Lin}, H., {Long}, D.~C., {Loomis}, C.~P., {Loveday}, J., {Lupton}, R.~H.,
  {Malanushenko}, O., {Malanushenko}, V., {Mandelbaum}, R., {Margon}, B.,
  {Marriner}, J.~P., {Mart{\'{\i}}nez-Delgado}, D., {Matsubara}, T., {McGehee},
  P.~M., {McKay}, T.~A., {Meiksin}, A., {Morrison}, H.~L., {Munn}, J.~A.,
  {Nakajima}, R., {Neilsen}, Jr., E.~H., {Newberg}, H.~J., {Nichol}, R.~C.,
  {Nicinski}, T., {Nieto-Santisteban}, M., {Nitta}, A., {Okamura}, S., {Owen},
  R., {Oyaizu}, H., {Padmanabhan}, N., {Pan}, K., {Park}, C., {Peoples}, J.~J.,
  {Pier}, J.~R., {Pope}, A.~C., {Purger}, N., {Raddick}, M.~J., {Re Fiorentin},
  P., {Richards}, G.~T., {Richmond}, M.~W., {Riess}, A.~G., {Rix}, H.-W.,
  {Rockosi}, C.~M., {Sako}, M., {Schlegel}, D.~J., {Schneider}, D.~P.,
  {Schreiber}, M.~R., {Schwope}, A.~D., {Seljak}, U., {Sesar}, B., {Sheldon},
  E., {Shimasaku}, K., {Sivarani}, T., {Smith}, J.~A., {Snedden}, S.~A.,
  {Steinmetz}, M., {Strauss}, M.~A., {SubbaRao}, M., {Suto}, Y., {Szalay},
  A.~S., {Szapudi}, I., {Szkody}, P., {Tegmark}, M., {Thakar}, A.~R.,
  {Tremonti}, C.~A., {Tucker}, D.~L., {Uomoto}, A., {Vanden Berk}, D.~E.,
  {Vandenberg}, J., {Vidrih}, S., {Vogeley}, M.~S., {Voges}, W., {Vogt}, N.~P.,
  {Wadadekar}, Y., {Weinberg}, D.~H., {West}, A.~A., {White}, S.~D.~M.,
  {Wilhite}, B.~C., {Yanny}, B., {Yocum}, D.~R., {York}, D.~G., {Zehavi}, I.,
  \& {Zucker}, D.~B. 2008, \apjs, 175, 297

\bibitem[{Arce \& Goodman(1999)}]{AG99}
Arce, H.~G., \& Goodman, A.~A. 1999, The Astrophysical Journal, 512, L135

\bibitem[{{Baldry} {et~al.}(2004){Baldry}, {Glazebrook}, {Brinkmann},
  {Ivezi{\'c}}, {Lupton}, {Nichol}, \& {Szalay}}]{baldry04}
{Baldry}, I.~K., {Glazebrook}, K., {Brinkmann}, J., {Ivezi{\'c}}, {\v Z}.,
  {Lupton}, R.~H., {Nichol}, R.~C., \& {Szalay}, A.~S. 2004, \apj, 600, 681

\bibitem[{{Balogh} {et~al.}(2004){Balogh}, {Baldry}, {Nichol}, {Miller},
  {Bower}, \& {Glazebrook}}]{balogh04}
{Balogh}, M.~L., {Baldry}, I.~K., {Nichol}, R., {Miller}, C., {Bower}, R., \&
  {Glazebrook}, K. 2004, \apjl, 615, L101

\bibitem[{Blanton \& Roweis(2007)}]{BR07}
Blanton, M.~R., \& Roweis, S. 2007, The Astronomical Journal, 133, 734

\bibitem[{{Blanton} {et~al.}(2005){Blanton}, {Schlegel}, {Strauss},
  {Brinkmann}, {Finkbeiner}, {Fukugita}, {Gunn}, {Hogg}, {Ivezi{\'c}}, {Knapp},
  {Lupton}, {Munn}, {Schneider}, {Tegmark}, \& {Zehavi}}]{blanton05-vagc}
{Blanton}, M.~R., {Schlegel}, D.~J., {Strauss}, M.~A., {Brinkmann}, J.,
  {Finkbeiner}, D., {Fukugita}, M., {Gunn}, J.~E., {Hogg}, D.~W., {Ivezi{\'c}},
  {\v Z}., {Knapp}, G.~R., {Lupton}, R.~H., {Munn}, J.~A., {Schneider}, D.~P.,
  {Tegmark}, M., \& {Zehavi}, I. 2005, \aj, 129, 2562

\bibitem[{Burstein \& Heiles(1978)}]{BH78}
Burstein, D., \& Heiles, C. 1978, Astrophysical Journal, 225, 40, a{\&}AA ID.
  AAA022.131.051

\bibitem[{Cambr{\'e}sy {et~al.}(2001)Cambr{\'e}sy, Boulanger, Lagache, \&
  Stepnik}]{Cambresy01}
Cambr{\'e}sy, L., Boulanger, F., Lagache, G., \& Stepnik, B. 2001, Astronomy
  and Astrophysics, 375, 999

\bibitem[{Chen {et~al.}(1999)Chen, Figueras, Torra, Jordi, Luri, \&
  Galad{\'\i}-Enr{\'\i}quez}]{Chen99}
Chen, B., Figueras, F., Torra, J., Jordi, C., Luri, X., \&
  Galad{\'\i}-Enr{\'\i}quez, D. 1999, Astronomy and Astrophysics, 352, 459

\bibitem[{Cooper {et~al.}(2008)Cooper, Newman, Weiner, Yan, Willmer, Bundy,
  Coil, Conselice, Davis, Faber, Gerke, Guhathakurta, Koo, \&
  Noeske}]{cooper08}
Cooper, M., Newman, J., Weiner, B., Yan, R., Willmer, C., Bundy, K., Coil, A.,
  Conselice, C., Davis, M., Faber, S., Gerke, B., Guhathakurta, P., Koo, D., \&
  Noeske, K. 2008, Monthly Notices of the Royal Astronomical Society, 383, 1058

\bibitem[{Doi {et~al.}(2010)Doi, Tanaka, Fukugita, Gunn, Yasuda, Ivezic,
  Brinkmann, de~Haars, Kleinman, Krzesinski, \& Leger}]{Doi10}
Doi, M., Tanaka, M., Fukugita, M., Gunn, J.~E., Yasuda, N., Ivezic, Z.,
  Brinkmann, J., de~Haars, E., Kleinman, S.~J., Krzesinski, J., \& Leger, R.~F.
  2010, arXiv, astro-ph.IM

\bibitem[{{Eisenstein} {et~al.}(2001){Eisenstein}, {Annis}, {Gunn}, {Szalay},
  {Connolly}, {Nichol}, {Bahcall}, {Bernardi}, {Burles}, {Castander},
  {Fukugita}, {Hogg}, {Ivezi{\'c}}, {Knapp}, {Lupton}, {Narayanan}, {Postman},
  {Reichart}, {Richmond}, {Schneider}, {Schlegel}, {Strauss}, {SubbaRao},
  {Tucker}, {Vanden Berk}, {Vogeley}, {Weinberg}, \& {Yanny}}]{eisenstein01}
{Eisenstein}, D.~J., {Annis}, J., {Gunn}, J.~E., {Szalay}, A.~S., {Connolly},
  A.~J., {Nichol}, R.~C., {Bahcall}, N.~A., {Bernardi}, M., {Burles}, S.,
  {Castander}, F.~J., {Fukugita}, M., {Hogg}, D.~W., {Ivezi{\'c}}, {\v Z}.,
  {Knapp}, G.~R., {Lupton}, R.~H., {Narayanan}, V., {Postman}, M., {Reichart},
  D.~E., {Richmond}, M., {Schneider}, D.~P., {Schlegel}, D.~J., {Strauss},
  M.~A., {SubbaRao}, M., {Tucker}, D.~L., {Vanden Berk}, D., {Vogeley}, M.~S.,
  {Weinberg}, D.~H., \& {Yanny}, B. 2001, \aj, 122, 2267

\bibitem[{{Graves} {et~al.}(2009){Graves}, {Faber}, \&
  {Schiavon}}]{graves09_paperI}
{Graves}, G.~J., {Faber}, S.~M., \& {Schiavon}, R.~P. 2009, \apj, 693, 486

\bibitem[{Gunn {et~al.}(1998)Gunn, Carr, Rockosi, Sekiguchi, Berry, Elms,
  de~Haas, Ivezi{\'c}, Knapp, Lupton, Pauls, Simcoe, Hirsch, Sanford, Wang,
  York, Harris, Annis, Bartozek, Boroski, Bakken, Haldeman, Kent, Holm,
  Holmgren, Petravick, Prosapio, Rechenmacher, Doi, Fukugita, Shimasaku, Okada,
  Hull, Siegmund, Mannery, Blouke, Heidtman, Schneider, Lucinio, \&
  Brinkman}]{Gunn98}
Gunn, J., Carr, M., Rockosi, C., Sekiguchi, M., Berry, K., Elms, B., de~Haas,
  E., Ivezi{\'c}, {\v Z}., Knapp, G., Lupton, R., Pauls, G., Simcoe, R.,
  Hirsch, R., Sanford, D., Wang, S., York, D., Harris, F., Annis, J., Bartozek,
  L., Boroski, W., Bakken, J., Haldeman, M., Kent, S., Holm, S., Holmgren, D.,
  Petravick, D., Prosapio, A., Rechenmacher, R., Doi, M., Fukugita, M.,
  Shimasaku, K., Okada, N., Hull, C., Siegmund, W., Mannery, E., Blouke, M.,
  Heidtman, D., Schneider, D., Lucinio, R., \& Brinkman, J. 1998, Astronomical
  Journal, 116, 3040

\bibitem[{{Hogg} {et~al.}(2003){Hogg}, {Blanton}, {Eisenstein}, {Gunn},
  {Schlegel}, {Zehavi}, {Bahcall}, {Brinkmann}, {Csabai}, {Schneider},
  {Weinberg}, \& {York}}]{hogg03}
{Hogg}, D.~W., {Blanton}, M.~R., {Eisenstein}, D.~J., {Gunn}, J.~E.,
  {Schlegel}, D.~J., {Zehavi}, I., {Bahcall}, N.~A., {Brinkmann}, J., {Csabai},
  I., {Schneider}, D.~P., {Weinberg}, D.~H., \& {York}, D.~G. 2003, \apjl, 585,
  L5

\bibitem[{Kalberla {et~al.}(2005)Kalberla, Burton, Hartmann, Arnal, Bajaja,
  Morras, \& P{\"o}ppel}]{Kalberla2005}
Kalberla, P.~M.~W., Burton, W.~B., Hartmann, D., Arnal, E.~M., Bajaja, E.,
  Morras, R., \& P{\"o}ppel, W.~G.~L. 2005, \aap, 440, 775

\bibitem[{{Knapp} \& {Kerr}(1974)}]{KK74}
{Knapp}, G.~R., \& {Kerr}, F.~J. 1974, \aap, 35, 361

\bibitem[{Markwardt(2009)}]{Markwardt09}
Markwardt, C.~B. 2009, Astronomical Data Analysis Software and Systems XVIII
  ASP Conference Series, 411, 251

\bibitem[{Padmanabhan {et~al.}(2008)Padmanabhan, Schlegel, Finkbeiner,
  Barentine, Blanton, Brewington, Gunn, Harvanek, Hogg, Ivezi{\'c}, Johnston,
  Kent, Kleinman, Knapp, Krzesinski, Long, Neilsen, Nitta, Loomis, Lupton,
  Roweis, Snedden, Strauss, \& Tucker}]{padmanabhan08}
Padmanabhan, N., Schlegel, D.~J., Finkbeiner, D.~P., Barentine, J.~C., Blanton,
  M.~R., Brewington, H.~J., Gunn, J.~E., Harvanek, M., Hogg, D.~W., Ivezi{\'c},
  {\v Z}., Johnston, D., Kent, S.~M., Kleinman, S.~J., Knapp, G.~R.,
  Krzesinski, J., Long, D., Neilsen, E.~H., Nitta, A., Loomis, C., Lupton,
  R.~H., Roweis, S., Snedden, S.~A., Strauss, M.~A., \& Tucker, D.~L. 2008, The
  Astrophysical Journal, 674, 1217

\bibitem[{Peek \& Heiles(2008)}]{PH08}
Peek, J. E.~G., \& Heiles, C. 2008, eprint arXiv, 0810, 1283, 25 pages, 10
  figures

\bibitem[{Rowles \& Froebrich(2009)}]{RF09}
Rowles, J., \& Froebrich, D. 2009, Monthly Notices of the Royal Astronomical
  Society, 395, 1640

\bibitem[{Schlegel {et~al.}(1998)Schlegel, Finkbeiner, \& Davis}]{SFD98}
Schlegel, D.~J., Finkbeiner, D.~P., \& Davis, M. 1998, \apj, 500, 525

\bibitem[{Shane \& Wirtanen(1967)}]{SW67}
Shane, C.~D., \& Wirtanen, C.~A. 1967, Publications of Lick Observatory

\bibitem[{Stanek(1998)}]{Stanek99}
Stanek, K.~Z. 1998, arXiv, astro-ph, submitted to the ApJ Letters, 10 pages, 3
  figures

\bibitem[{{Strateva} {et~al.}(2001){Strateva}, {Ivezi{\'c}}, {Knapp},
  {Narayanan}, {Strauss}, {Gunn}, {Lupton}, {Schlegel}, {Bahcall}, {Brinkmann},
  {Brunner}, {Budav{\'a}ri}, {Csabai}, {Castander}, {Doi}, {Fukugita}, {Gy{\H
  o}ry}, {Hamabe}, {Hennessy}, {Ichikawa}, {Kunszt}, {Lamb}, {McKay},
  {Okamura}, {Racusin}, {Sekiguchi}, {Schneider}, {Shimasaku}, \&
  {York}}]{strateva01}
{Strateva}, I., {Ivezi{\'c}}, {\v Z}., {Knapp}, G.~R., {Narayanan}, V.~K.,
  {Strauss}, M.~A., {Gunn}, J.~E., {Lupton}, R.~H., {Schlegel}, D., {Bahcall},
  N.~A., {Brinkmann}, J., {Brunner}, R.~J., {Budav{\'a}ri}, T., {Csabai}, I.,
  {Castander}, F.~J., {Doi}, M., {Fukugita}, M., {Gy{\H o}ry}, Z., {Hamabe},
  M., {Hennessy}, G., {Ichikawa}, T., {Kunszt}, P.~Z., {Lamb}, D.~Q., {McKay},
  T.~A., {Okamura}, S., {Racusin}, J., {Sekiguchi}, M., {Schneider}, D.~P.,
  {Shimasaku}, K., \& {York}, D. 2001, \aj, 122, 1861

\bibitem[{{Strauss} {et~al.}(2002){Strauss}, {Weinberg}, {Lupton}, {Narayanan},
  {Annis}, {Bernardi}, {Blanton}, {Burles}, {Connolly}, {Dalcanton}, {Doi},
  {Eisenstein}, {Frieman}, {Fukugita}, {Gunn}, {Ivezi{\'c}}, {Kent}, {Kim},
  {Knapp}, {Kron}, {Munn}, {Newberg}, {Nichol}, {Okamura}, {Quinn}, {Richmond},
  {Schlegel}, {Shimasaku}, {SubbaRao}, {Szalay}, {Vanden Berk}, {Vogeley},
  {Yanny}, {Yasuda}, {York}, \& {Zehavi}}]{strauss02}
{Strauss}, M.~A., {Weinberg}, D.~H., {Lupton}, R.~H., {Narayanan}, V.~K.,
  {Annis}, J., {Bernardi}, M., {Blanton}, M., {Burles}, S., {Connolly}, A.~J.,
  {Dalcanton}, J., {Doi}, M., {Eisenstein}, D., {Frieman}, J.~A., {Fukugita},
  M., {Gunn}, J.~E., {Ivezi{\'c}}, {\v Z}., {Kent}, S., {Kim}, R.~S.~J.,
  {Knapp}, G.~R., {Kron}, R.~G., {Munn}, J.~A., {Newberg}, H.~J., {Nichol},
  R.~C., {Okamura}, S., {Quinn}, T.~R., {Richmond}, M.~W., {Schlegel}, D.~J.,
  {Shimasaku}, K., {SubbaRao}, M., {Szalay}, A.~S., {Vanden Berk}, D.,
  {Vogeley}, M.~S., {Yanny}, B., {Yasuda}, N., {York}, D.~G., \& {Zehavi}, I.
  2002, \aj, 124, 1810

\bibitem[{{Yan} {et~al.}(2006){Yan}, {Newman}, {Faber}, {Konidaris}, {Koo}, \&
  {Davis}}]{yan06}
{Yan}, R., {Newman}, J.~A., {Faber}, S.~M., {Konidaris}, N., {Koo}, D., \&
  {Davis}, M. 2006, \apj, 648, 281

\bibitem[{Yasuda {et~al.}(2007)Yasuda, Fukugita, \& Schneider}]{YFS07}
Yasuda, N., Fukugita, M., \& Schneider, D.~P. 2007, The Astronomical Journal,
  134, 698

\bibitem[{{York} {et~al.}(2000){York}, {Adelman}, {Anderson}, {Anderson},
  {Annis}, {Bahcall}, {Bakken}, {Barkhouser}, {Bastian}, {Berman}, {Boroski},
  {Bracker}, {Briegel}, {Briggs}, {Brinkmann}, {Brunner}, {Burles}, {Carey},
  {Carr}, {Castander}, {Chen}, {Colestock}, {Connolly}, {Crocker}, {Csabai},
  {Czarapata}, {Davis}, {Doi}, {Dombeck}, {Eisenstein}, {Ellman}, {Elms},
  {Evans}, {Fan}, {Federwitz}, {Fiscelli}, {Friedman}, {Frieman}, {Fukugita},
  {Gillespie}, {Gunn}, {Gurbani}, {de Haas}, {Haldeman}, {Harris}, {Hayes},
  {Heckman}, {Hennessy}, {Hindsley}, {Holm}, {Holmgren}, {Huang}, {Hull},
  {Husby}, {Ichikawa}, {Ichikawa}, {Ivezi{\'c}}, {Kent}, {Kim}, {Kinney},
  {Klaene}, {Kleinman}, {Kleinman}, {Knapp}, {Korienek}, {Kron}, {Kunszt},
  {Lamb}, {Lee}, {Leger}, {Limmongkol}, {Lindenmeyer}, {Long}, {Loomis},
  {Loveday}, {Lucinio}, {Lupton}, {MacKinnon}, {Mannery}, {Mantsch}, {Margon},
  {McGehee}, {McKay}, {Meiksin}, {Merelli}, {Monet}, {Munn}, {Narayanan},
  {Nash}, {Neilsen}, {Neswold}, {Newberg}, {Nichol}, {Nicinski}, {Nonino},
  {Okada}, {Okamura}, {Ostriker}, {Owen}, {Pauls}, {Peoples}, {Peterson},
  {Petravick}, {Pier}, {Pope}, {Pordes}, {Prosapio}, {Rechenmacher}, {Quinn},
  {Richards}, {Richmond}, {Rivetta}, {Rockosi}, {Ruthmansdorfer}, {Sandford},
  {Schlegel}, {Schneider}, {Sekiguchi}, {Sergey}, {Shimasaku}, {Siegmund},
  {Smee}, {Smith}, {Snedden}, {Stone}, {Stoughton}, {Strauss}, {Stubbs},
  {SubbaRao}, {Szalay}, {Szapudi}, {Szokoly}, {Thakar}, {Tremonti}, {Tucker},
  {Uomoto}, {Vanden Berk}, {Vogeley}, {Waddell}, {Wang}, {Watanabe},
  {Weinberg}, {Yanny}, \& {Yasuda}}]{york00}
{York}, D.~G., {Adelman}, J., {Anderson}, Jr., J.~E., {Anderson}, S.~F.,
  {Annis}, J., {Bahcall}, N.~A., {Bakken}, J.~A., {Barkhouser}, R., {Bastian},
  S., {Berman}, E., {Boroski}, W.~N., {Bracker}, S., {Briegel}, C., {Briggs},
  J.~W., {Brinkmann}, J., {Brunner}, R., {Burles}, S., {Carey}, L., {Carr},
  M.~A., {Castander}, F.~J., {Chen}, B., {Colestock}, P.~L., {Connolly}, A.~J.,
  {Crocker}, J.~H., {Csabai}, I., {Czarapata}, P.~C., {Davis}, J.~E., {Doi},
  M., {Dombeck}, T., {Eisenstein}, D., {Ellman}, N., {Elms}, B.~R., {Evans},
  M.~L., {Fan}, X., {Federwitz}, G.~R., {Fiscelli}, L., {Friedman}, S.,
  {Frieman}, J.~A., {Fukugita}, M., {Gillespie}, B., {Gunn}, J.~E., {Gurbani},
  V.~K., {de Haas}, E., {Haldeman}, M., {Harris}, F.~H., {Hayes}, J.,
  {Heckman}, T.~M., {Hennessy}, G.~S., {Hindsley}, R.~B., {Holm}, S.,
  {Holmgren}, D.~J., {Huang}, C.-h., {Hull}, C., {Husby}, D., {Ichikawa},
  S.-I., {Ichikawa}, T., {Ivezi{\'c}}, {\v Z}., {Kent}, S., {Kim}, R.~S.~J.,
  {Kinney}, E., {Klaene}, M., {Kleinman}, A.~N., {Kleinman}, S., {Knapp},
  G.~R., {Korienek}, J., {Kron}, R.~G., {Kunszt}, P.~Z., {Lamb}, D.~Q., {Lee},
  B., {Leger}, R.~F., {Limmongkol}, S., {Lindenmeyer}, C., {Long}, D.~C.,
  {Loomis}, C., {Loveday}, J., {Lucinio}, R., {Lupton}, R.~H., {MacKinnon}, B.,
  {Mannery}, E.~J., {Mantsch}, P.~M., {Margon}, B., {McGehee}, P., {McKay},
  T.~A., {Meiksin}, A., {Merelli}, A., {Monet}, D.~G., {Munn}, J.~A.,
  {Narayanan}, V.~K., {Nash}, T., {Neilsen}, E., {Neswold}, R., {Newberg},
  H.~J., {Nichol}, R.~C., {Nicinski}, T., {Nonino}, M., {Okada}, N., {Okamura},
  S., {Ostriker}, J.~P., {Owen}, R., {Pauls}, A.~G., {Peoples}, J., {Peterson},
  R.~L., {Petravick}, D., {Pier}, J.~R., {Pope}, A., {Pordes}, R., {Prosapio},
  A., {Rechenmacher}, R., {Quinn}, T.~R., {Richards}, G.~T., {Richmond}, M.~W.,
  {Rivetta}, C.~H., {Rockosi}, C.~M., {Ruthmansdorfer}, K., {Sandford}, D.,
  {Schlegel}, D.~J., {Schneider}, D.~P., {Sekiguchi}, M., {Sergey}, G.,
  {Shimasaku}, K., {Siegmund}, W.~A., {Smee}, S., {Smith}, J.~A., {Snedden},
  S., {Stone}, R., {Stoughton}, C., {Strauss}, M.~A., {Stubbs}, C., {SubbaRao},
  M., {Szalay}, A.~S., {Szapudi}, I., {Szokoly}, G.~P., {Thakar}, A.~R.,
  {Tremonti}, C., {Tucker}, D.~L., {Uomoto}, A., {Vanden Berk}, D., {Vogeley},
  M.~S., {Waddell}, P., {Wang}, S.-i., {Watanabe}, M., {Weinberg}, D.~H.,
  {Yanny}, B., \& {Yasuda}, N. 2000, \aj, 120, 1579

\end{thebibliography}

\end{document}